\documentclass[aps,prc,preprint,groupedaddress,showpacs]{revtex4-1}

\usepackage{graphicx}
\usepackage{bm}

\begin{document}


\title{Modified parameter-sets
of M3Y-type semi-realistic nucleon-nucleon interaction
for nuclear structure studies}


\author{H. Nakada}
\email[E-mail:\,\,]{nakada@faculty.chiba-u.jp}
\affiliation{Department of Physics, Graduate School of Science,
 Chiba University\\
Yayoi-cho 1-33, Inage, Chiba 263-8522, Japan}


\date{\today}

\begin{abstract}
New parameter-sets of the M3Y-type semi-realistic
nucleon-nucleon interaction are obtained,
by taking into account the whole contribution of the interaction
to the pairing.
The interactions are applicable to nuclear structure studies
via mean-field calculations and their extensions.
Implementing self-consistent mean-field calculations
for the spherical nuclei,
we confirm that the results do not change significantly
from the previous ones,
in which the parameters were determined
by partly discarding influence of the density-dependent repulsion
on the pairing.
\end{abstract}


\maketitle



\section{Introduction\label{sec:intro}}

Mean-field (MF) theories provide us with a basic tool
to study nuclear structure
from nucleonic degrees of freedom.
However, it is yet difficult to reproduce
structure of medium- to heavy-mass nuclei
with the bare nucleon-nucleon ($NN$) interaction
to good accuracy,
and most of the MF calculations have been performed
with phenomenologically determined effective interactions
(or energy density functionals).

The Michigan 3-range Yukawa (M3Y) interaction~\cite{ref:M3Y}
was obtained by fitting the Yukawa functions
to Brueckner's $G$-matrix.
As the crucial role of the chiral symmetry breaking has been recognized,
it is desirable to take into account the pionic effects
in the $NN$ interaction.
The M3Y interaction contains the central part
of the one-pion exchange potential (OPEP)
and the tensor part that is corrected due to the medium effects
as well as to the $\rho$-meson exchange.
The author has developed \textit{semi-realistic}
effective $NN$ interactions~\cite{ref:Nak03,ref:Nak08b},
by modifying the M3Y interaction
so as to reproduce the saturation properties and the $\ell s$ splitting
while keeping the central part of the OPEP.
The tensor channels have been pointed out to be significant
in the $Z$- or $N$-dependence of the shell structure~\cite{ref:Vtn}.
We have obtained parameter-sets
in which the tensor channels are maintained (\textit{e.g.} the set M3Y-P5),
and one in which the tensor channels are dropped (M3Y-P4)~\cite{ref:Nak08b}.
By implementing MF calculations, it has been shown
that semi-realistic interactions could give different shell structure
from the widely used Skyrme and Gogny interactions,
and that of the semi-realistic interaction
including the tensor channels is often favorable
if compared to the experimental data.
Significance of the tensor channels in the magnetic excitations
has also been confirmed~\cite{ref:Shi08}
via calculations in the random-phase approximation.

The parameter-sets developed in Ref.~\cite{ref:Nak08b}
were obtained by fitting them to pairing properties,
as well as to data of doubly magic nuclei.
However, in the calculations of finite nuclei in Ref.~\cite{ref:Nak08b},
influence of the density-dependent contact force
on the pairing was not fully taken into account.
Because we find that this influence is not negligibly small,
we propose modified parameter-sets in this report,
which will be useful in future studies of nuclear structure.
The new parameter-sets are refitted to the same quantities
as the old ones, as will be described in the subsequent section.

\section{M3Y-type interaction\label{sec:M3Y}}

We consider the effective $NN$ interaction that has the following form,
\begin{eqnarray} v_{ij} &=& v_{ij}^{(\mathrm{C})}
 + v_{ij}^{(\mathrm{LS})} + v_{ij}^{(\mathrm{TN})}
 + v_{ij}^{(\mathrm{DD})}\,;\nonumber\\
v_{ij}^{(\mathrm{C})} &=& \sum_n \big(t_n^{(\mathrm{SE})} P_\mathrm{SE}
+ t_n^{(\mathrm{TE})} P_\mathrm{TE} + t_n^{(\mathrm{SO})} P_\mathrm{SO}
+ t_n^{(\mathrm{TO})} P_\mathrm{TO}\big)
 f_n^{(\mathrm{C})} (r_{ij})\,,\nonumber\\
v_{ij}^{(\mathrm{DD})} &=& \big(t_\rho^{(\mathrm{SE})} P_\mathrm{SE}\cdot
 [\rho(\mathbf{r}_i)]^{\alpha^{(\mathrm{SE})}}
 + t_\rho^{(\mathrm{TE})} P_\mathrm{TE}\cdot
 [\rho(\mathbf{r}_i)]^{\alpha^{(\mathrm{TE})}}\big)
 \,\delta(\mathbf{r}_{ij})\,,
\label{eq:effint}\end{eqnarray}
where $\mathbf{r}_{ij}= \mathbf{r}_i - \mathbf{r}_j$,
$r_{ij}=|\mathbf{r}_{ij}|$
with $i$ and $j$ representing the indices of nucleons,
and $\rho(\mathbf{r})$ denotes the nucleon density.
$P_\mathrm{SE}$, $P_\mathrm{TE}$, $P_\mathrm{SO}$ and $P_\mathrm{TO}$
denote the projection operators
on the singlet-even (SE), triplet-even (TE), singlet-odd (SO)
and triplet-odd (TO) two-particle states.
The density-dependent contact force $v_{ij}^{(\mathrm{DD})}$
has been added to reproduce the saturation~\cite{ref:Nak03}.
The Yukawa function $f_n^{(\mathrm{X})}(r)
=e^{-\mu_n^{(\mathrm{X})} r}/\mu_n^{(\mathrm{X})} r$
is assumed for $\mathrm{X}=\mathrm{C}$, $\mathrm{LS}$
and $\mathrm{TN}$.
The modified parameter-sets M3Y-P4$'$ and P5$'$
are presented in Table~\ref{tab:param_M3Y},
in comparison with the previous sets M3Y-P4 and P5.
The non-central channels $v_{ij}^{(\mathrm{LS})}$
and $v_{ij}^{(\mathrm{TN})}$ are not changed;
$v_{ij}^{(\mathrm{LS})}$ and $v_{ij}^{(\mathrm{TN})}$
of M3Y-P4$'$ (M3Y-P5$'$) are identical to
those of M3Y-P4 (M3Y-P5)~\cite{ref:Nak08b}.
The M3Y-P5$'$ interaction has $v_{ij}^{(\mathrm{TN})}$
determined from the $G$-matrix~\cite{ref:M3Y-P},
while $v_{ij}^{(\mathrm{TN})}=0$ is assumed in M3Y-P4$'$.
Not displayed in Table~\ref{tab:param_M3Y},
the longest range ($n=3$) part of $v_{ij}^{(\mathrm{C})}$
is kept to be $v^{(\mathrm{C})}_\mathrm{OPEP}$
(the central channels of the OPEP).
The other strength parameters are fitted
to the measured binding energies of $^{16}$O and $^{208}$Pb
and to the even-odd mass differences of the Sn isotopes,
in the Hartree-Fock (HF)
or the Hartree-Fock-Bogolyubov (HFB) approximation.
Compared to the old sets M3Y-P4 and P5,
in M3Y-P4$'$ and P5$'$ the SE channel is more attractive
below the saturation density,
and its influence on the saturation and the symmetry energies
is compensated by the other central channels.

\begin{table}
\begin{center}
\caption{Parameters of M3Y-type interactions.
\label{tab:param_M3Y}}
\begin{tabular}{ccr@{.}lr@{.}lr@{.}lr@{.}l}
\hline\hline
parameters &&
 \multicolumn{2}{c}{~~M3Y-P4~~} & \multicolumn{2}{c}{~~M3Y-P4$'$\,~} &
 \multicolumn{2}{c}{~~M3Y-P5~~} & \multicolumn{2}{c}{~~M3Y-P5$'$\,~} \\
 \hline
$1/\mu_1^{(\mathrm{C})}$ &(fm)&
 $0$&$25$ & $0$&$25$ & $0$&$25$ & $0$&$25$ \\
$t_1^{(\mathrm{SE})}$ &(MeV)& $8027$& & $8027$& & $8027$& & $8027$& \\
$t_1^{(\mathrm{TE})}$ &(MeV)& $5503$& & $5671$& & $5576$& & $5576$& \\
$t_1^{(\mathrm{SO})}$ &(MeV)& $-12000$& & $-12500$& & $-1418$& & $-1418$& \\
$t_1^{(\mathrm{TO})}$ &(MeV)& $3700$& & $4200$& & $11345$& & $11345$& \\
$1/\mu_2^{(\mathrm{C})}$ &(fm)&
 $0$&$40$ & $0$&$40$ & $0$&$40$ & $0$&$40$ \\
$t_2^{(\mathrm{SE})}$ &(MeV)& $-2637$& &$-2765$& & $-2650$& & $-2760$& \\
$t_2^{(\mathrm{TE})}$ &(MeV)& $-4183$& & $-4118$& & $-4170$& & $-4029$& \\
$t_2^{(\mathrm{SO})}$ &(MeV)& $4500$& & $3000$& & $2880$& & $1700$& \\
$t_2^{(\mathrm{TO})}$ &(MeV)& $-1000$& & $-800$& & $-1780$& & $-1715$& \\
$\alpha^{(\mathrm{SE})}$ &&
 \multicolumn{2}{c}{$1$} & \multicolumn{2}{c}{$1$} &
 \multicolumn{2}{c}{$1$} & \multicolumn{2}{c}{$1$} \\
$t_\rho^{(\mathrm{SE})}$ &(MeV$\cdot$fm$^3$)
 & $248$& & $180$& & $126$& & $178$& \\
$\alpha^{(\mathrm{TE})}$ &&
 \multicolumn{2}{c}{$1/3$} & \multicolumn{2}{c}{$1/3$} &
 \multicolumn{2}{c}{$1/3$} & \multicolumn{2}{c}{$1/3$} \\
$t_\rho^{(\mathrm{TE})}$ &(MeV$\cdot$fm)
 & $1142$& & $1155$&  & $1147$& & $1128$& \\
\hline\hline
\end{tabular}
\end{center}
\end{table}

\section{Nuclear matter properties\label{sec:NMprop}}

Energy of the nuclear matter is a function of
$\rho = \sum_{\sigma\tau} \rho_{\tau\sigma}$,
$\eta_s = \sum_{\sigma\tau}\sigma\rho_{\tau\sigma}\big/\rho$,
$\eta_t = \sum_{\sigma\tau}\tau\rho_{\tau\sigma}\big/\rho$ and
$\eta_{st} = \sum_{\sigma\tau}\sigma\tau\rho_{\tau\sigma}\big/\rho$
in the HF approximation,
with $\rho_{\tau\sigma}$ ($\tau=p,n$ and $\sigma=\uparrow,\downarrow$)
stands for densities depending on the spin and the isospin.
Formulas to calculate the nuclear matter energy and its derivatives
for given $\rho_{\tau\sigma}$ have been derived
in Ref.~\cite{ref:Nak03}.

The energy per nucleon $\mathcal{E}=E/A$
has its minimum  $\mathcal{E}_0$,
at the saturation density $\rho_0$ (equivalently, $k_{\mathrm{F}0}$).
The incompressibility and the volume symmetry energy
are defined by
\begin{equation}
 \mathcal{K} = 9\rho^2 \left.\frac{\partial^2\mathcal{E}}{\partial\rho^2}
       \right\vert_0\,,\quad
 a_t = \left. \frac{1}{2} \frac{\partial^2\mathcal{E}}{\partial\eta_t^2}
	\right\vert_0\,.
\end{equation}
Here $~\vert_0$ indicates evaluation at the saturation point.
The curvatures with respect to $\eta_s$ and $\eta_{st}$ are
denoted by $a_s$ and $a_{st}$.
The $k$-mass $M_0^\ast$ is defined by a derivative of the s.p. energy.
Density-dependence of the symmetry energy is
typically represented by
\begin{equation} \mathcal{L}_t = \left.\frac{3}{2}\rho
 \frac{\partial^3\mathcal{E}}{\partial\rho\,\partial\eta_t^2}
  \right\vert_0\,.
\end{equation}
These quantities calculated from the new semi-realistic interactions
are tabulated in Table~\ref{tab:NMsat}.
For comparison,
the values obtained by the D1S~\cite{ref:D1S} and D1M~\cite{ref:D1M}
parameter-set of the Gogny interaction are also displayed.
We also present the Landau-Migdal (LM) parameters in Table~\ref{tab:LM},
which are calculated by using the formulas given in Ref.~\cite{ref:Nak03}.
See Ref.~\cite{ref:Nak03} also for definition of the LM parameters.
It has been known that $g'_0\approx 1$~\cite{ref:g'0},
which is reproduced by the semi-realistic M3Y-type interactions
primarily due to $v^{(\mathrm{C})}_\mathrm{OPEP}$.

\begin{table}
\begin{center}
\caption{Nuclear matter properties at the saturation point.
\label{tab:NMsat}}
\begin{tabular}{ccrrrr}
\hline\hline
&&~~~~D1S~~ &~~~~D1M~~ &~~M3Y-P4$'$ &~~M3Y-P5$'$ \\ \hline
$k_{\mathrm{F}0}$ & (fm$^{-1}$) & $1.342$~~& $1.346$~~& $1.340$~~& $1.340$~~\\
$\mathcal{E}_0$ & (MeV) & $-16.01$~~& $-16.02$~~& $-16.05$~~& $-16.14$~~\\
$\mathcal{K}$ & (MeV) & $202.9$~~& $225.0$~~& $230.4$~~& $239.1$~~\\
$M^\ast_0/M$ && $0.697$~~& $0.746$~~& $0.653$~~& $0.637$~~\\
$a_t$ & (MeV) & $31.12$~~& $28.55$~~& $28.49$~~& $28.42$~~\\
$a_s$ & (MeV) & $26.18$~~& $16.56$~~& $23.35$~~& $23.70$~~\\
$a_{st}$ & (MeV) & $29.13$~~& $28.71$~~& $38.66$~~& $39.11$~~\\
$\mathcal{L}_t$ & (MeV) & $22.44$~~& $24.83$~~& $21.15$~~& $25.12$~~\\
\hline\hline
\end{tabular}
\end{center}
\end{table}

\begin{table}
\begin{center}
\caption{Landau-Migdal parameters at the saturation point.
\label{tab:LM}}
\begin{tabular}{cr@{.}lr@{.}lr@{.}lr@{.}lr@{.}lr@{.}l}
\hline\hline
\hspace*{1cm} & \multicolumn{2}{c}{~~D1S~~~~} & \multicolumn{2}{c}{~~D1M~~~~} &
 \multicolumn{2}{c}{M3Y-P4$'$\,~} & \multicolumn{2}{c}{M3Y-P5$'$\,~} \\ \hline
$f_0$ & $-0$&$369$ & $-0$&$255$ & $-0$&$327$ & $-0$&$318$ \\
$f_1$ & $-0$&$909$ & $-0$&$762$ & $-1$&$042$ & $-1$&$089$ \\
$f_2$ & $-0$&$558$ & $-0$&$302$ & $-0$&$431$ & $-0$&$363$ \\
$f_3$ & $-0$&$157$ & $-0$&$058$ & $-0$&$208$ & $-0$&$182$ \\
 \hline
$f'_0$ & $0$&$743$ & $0$&$701$ & $0$&$499$ & $0$&$458$ \\
$f'_1$ & $0$&$470$ & $0$&$378$ & $0$&$631$ & $0$&$600$ \\
$f'_2$ & $0$&$342$ & $0$&$633$ & $0$&$245$ & $0$&$245$ \\
$f'_3$ & $0$&$100$ & $0$&$137$ & $0$&$096$ & $0$&$096$ \\
 \hline
$g_0$ & $0$&$466$ & $-0$&$013$ & $0$&$228$ & $0$&$216$ \\
$g_1$ & $-0$&$184$ & $-0$&$380$ & $0$&$263$ & $0$&$255$ \\
$g_2$ & $0$&$245$ & $0$&$483$ & $0$&$162$ & $0$&$168$ \\
$g_3$ & $0$&$091$ & $0$&$114$ & $0$&$078$ & $0$&$080$ \\
 \hline
$g'_0$ & $0$&$631$ & $0$&$711$ & $1$&$033$ & $1$&$007$ \\
$g'_1$ & $0$&$610$ & $0$&$652$ & $0$&$180$ & $0$&$146$ \\
$g'_2$ & $-0$&$038$ & $-0$&$243$ & $0$&$034$ & $0$&$044$ \\
$g'_3$ & $-0$&$036$ & $-0$&$064$ & $-0$&$004$ & $0$&$008$ \\
\hline\hline
\end{tabular}
\end{center}
\end{table}

\begin{figure}
\includegraphics[scale=0.9]{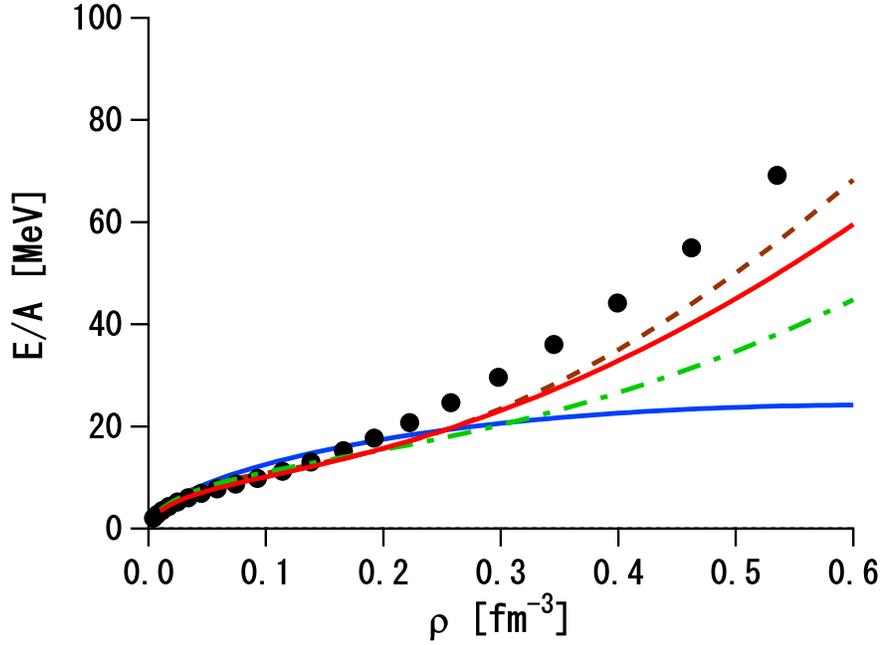}
\caption{Energy per nucleon $\mathcal{E}=E/A$ in the neutron matter.
Blue solid, brown dotted, green dot-dashed and red solid lines
represent the results with the D1S, D1M, M3Y-P4$'$ and P5$'$
interactions, respectively.
Circles are the result of Ref.~\protect\cite{ref:FP81}.
\label{fig:NME_M3Yc}}
\end{figure}

For the symmetric nuclear matter, M3Y-P4$'$ and P5$'$
give $\mathcal{E}(\rho)$ very close to each other,
and also to those of M3Y-P4 and P5~\cite{ref:Nak08b}.
$\mathcal{E}(\rho)$ in the neutron matter
is shown in Fig.~\ref{fig:NME_M3Yc}.
As in M3Y-P4 and P5,
the $\rho$-dependence of the SE channel in M3Y-P4$'$ and P5$'$
gives sizable repulsion at high $\rho$,
by which $\mathcal{E}(\rho)$ is qualitatively similar
to the result of Ref.~\cite{ref:FP81},
unlike the D1S case.

\section{Application to finite nuclei\label{sec:DMprop}}

The HF and HFB calculations are implemented in finite nuclei
by using the Gaussian expansion method~\cite{ref:NS02,ref:Nak06,
ref:Nak08,ref:NMYM09}
and adopting the Hamiltonian $H=H_N+V_C-H_\mathrm{c.m.}$,
where $H_N$, $V_C$ and $H_\mathrm{c.m.}$ represent
the effective nuclear Hamiltonian, the Coulomb interaction
and the center-of-mass Hamiltonian.
The exchange term of $V_C$ is treated exactly.
Both the one- and the two-body terms of $H_\mathrm{c.m.}$
are subtracted before iteration.

The binding energies and rms matter radii
of several doubly magic nuclei
are calculated in the spherical HF approximation.
The results of the new semi-realistic interactions are
compared with those of D1S and D1M,
as well as with the experimental data,
in Table~\ref{tab:DMprop}.
The new interactions reproduce the binding energies
and the rms matter radii
with similar accuracy to M3Y-P4 and P5.
The deviation in the binding energies
is $\lesssim 10\,\mathrm{MeV}$.
Though slightly worse than D1S and D1M,
we do not take this deviation seriously,
until correlations due to the residual interaction are taken into account.

\begin{table}
\begin{center}
\caption{Binding energies and rms matter radii
 of several doubly magic nuclei.
 Experimental data are taken
 from Refs.~\protect\cite{ref:mass,ref:O16-rad,ref:O24-rad,ref:rad}.
\label{tab:DMprop}}
\begin{tabular}{cccrrrrr}
\hline\hline
&&&~~~Exp.~~&~~~~~D1S~~&~~~~~D1M~~&~~M3Y-P4$'$&~~M3Y-P5$'$\\ \hline
$^{16}$O & $-E$ &(MeV)&
 $127.6$ & $129.5$ & $128.2$ & $127.8$ & $124.1$ \\
& $\sqrt{\langle r^2\rangle}$ &(fm)&
 $2.61$ & $2.61$ & $2.57$ & $2.59$ & $2.60$ \\
$^{24}$O & $-E$ &(MeV)&
 $168.5$ & $168.6$ & $167.3$ & $166.6$ & $166.4$ \\
& $\sqrt{\langle r^2\rangle}$ &(fm)&
 $3.19$ & $3.01$ & $2.98$ & $3.02$ & $3.02$ \\
$^{40}$Ca & $-E$ &(MeV)&
 $342.1$ & $344.6$ & $342.2$ & $338.7$ & $331.7$ \\
& $\sqrt{\langle r^2\rangle}$ &(fm)&
 $3.47$ & $3.37$ & $3.33$ & $3.36$ & $3.37$ \\
$^{48}$Ca & $-E$ &(MeV)&
 $416.0$ & $416.8$ & $414.6$ & $412.0$ & $411.5$ \\
& $\sqrt{\langle r^2\rangle}$ &(fm)&
 $3.57$ & $3.51$ & $3.48$ & $3.51$ & $3.51$ \\
$^{90}$Zr & $-E$ &(MeV)&
 $783.9$ & $785.9$ & $782.1$ & $776.9$ & $775.7$ \\
& $\sqrt{\langle r^2\rangle}$ &(fm)&
 $4.32$ & $4.24$ & $4.20$ & $4.23$ & $4.23$ \\
$^{132}$Sn & $-E$ &(MeV)&
 $1102.9$ & $1104.1$ & $1104.5$ & $1100.8$ & $1100.6$ \\
& $\sqrt{\langle r^2\rangle}$ &(fm)&
 --- & $4.77$ & $4.72$ & $4.77$ & $4.76$ \\
$^{208}$Pb & $-E$ &(MeV)&
 $1636.4$ & $1639.0$ & $1638.9$ & $1634.7$ & $1635.7$ \\
& $\sqrt{\langle r^2\rangle}$ &(fm)&
 $5.49$ & $5.51$ & $5.47$ & $5.51$ & $5.51$ \\
\hline\hline
\end{tabular}
\end{center}
\end{table}

\begin{figure}
\includegraphics[scale=1.0]{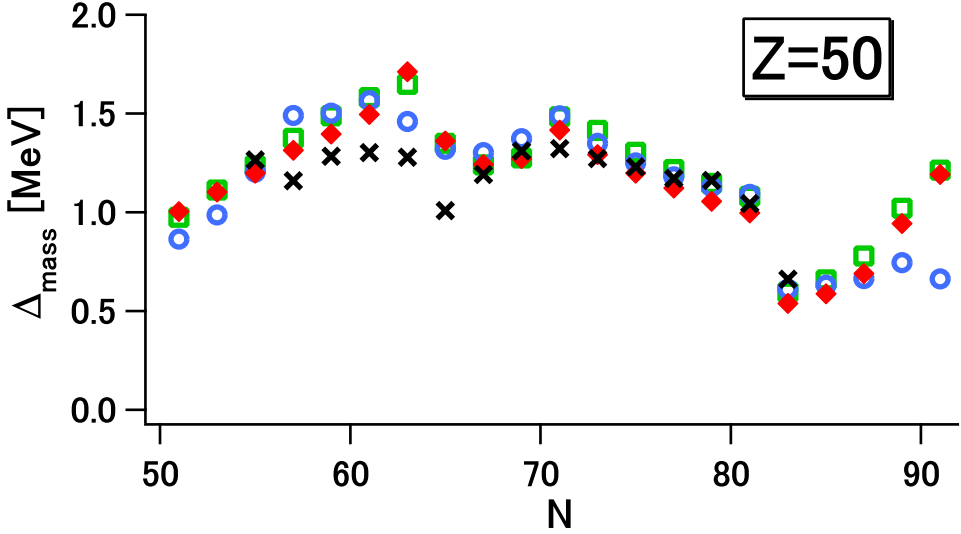}
\caption{Even-odd mass difference in the Sn isotopes,
$\Delta_\mathrm{mass}^{Z=50}(N)$.
The results of D1S, M3Y-P4$'$ and M3Y-P5$'$ are shown
by blue open circles, green open squares and red diamonds, respectively.
Experimental values, presented by black crosses,
are taken from Ref.~\protect\cite{ref:mass}.
\label{fig:Sn_Dmass}}
\end{figure}

In M3Y-P4$'$ and P5$'$,
some of the parameters in the SE channel are fitted
to the even-odd mass differences
$\Delta_\mathrm{mass}^Z(N)=E(Z,N)-\frac{1}{2}\big[E(Z,N+1)+E(Z,N-1)\big]$
($N=\mathrm{odd}$) of the $66\leq N\leq 80$ Sn isotopes,
in the spherical HFB approximation
with the $\ell\leq 7$ basis truncation.
Contribution of $v_{ij}^{(\mathrm{DD})}$ to the pairing
is fully taken into account.
The equal filling approximation~\cite{ref:EFA} is applied
to the odd-mass nuclei.
See arguments in Ref.~\cite{ref:Nak08b}
on influence of the $\ell$ truncation, the particle-number conservation
and the non-spherical mean fields.
$\Delta_\mathrm{mass}^{Z=50}(N)$ calculated by M3Y-P4$'$ and P5$'$
are compared with the experimental data and with those obtained by D1S
in Fig.~\ref{fig:Sn_Dmass}.
Though not shown,
D1M gives $\Delta_\mathrm{mass}^{Z=50}(N)$ quite close to that of D1S
except at $N=92$.
The mass difference at $N\sim 64$ and $90$ reflects
interaction-dependence of the shell structure.
In the M3Y-P4$'$ and P5$'$ interactions
the $N=64$ subshell effect is stronger than in D1S.
At $N\sim 90$ the M3Y-type interactions
yield larger mass difference than D1S,
because $n1f_{7/2}$ and $n2p_{3/2}$ well mix due to the pairing.

\begin{figure}
\includegraphics[scale=1.0]{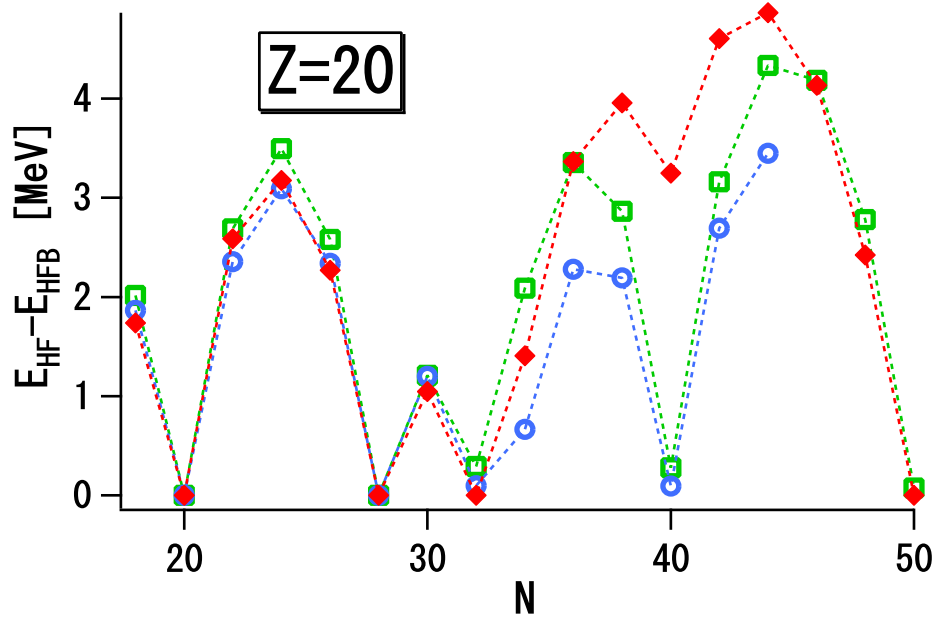}
\caption{Difference between the HF and HFB energies
for the Ca isotopes ($N=\mbox{even}$),
obtained from D1S, M3Y-P4$'$ and P5$'$.
See Fig.~\protect\ref{fig:Sn_Dmass} for conventions.
Dotted lines are drawn to guide eyes.
\label{fig:Z20_Epr}}
\end{figure}

\begin{figure}
\includegraphics[scale=1.0]{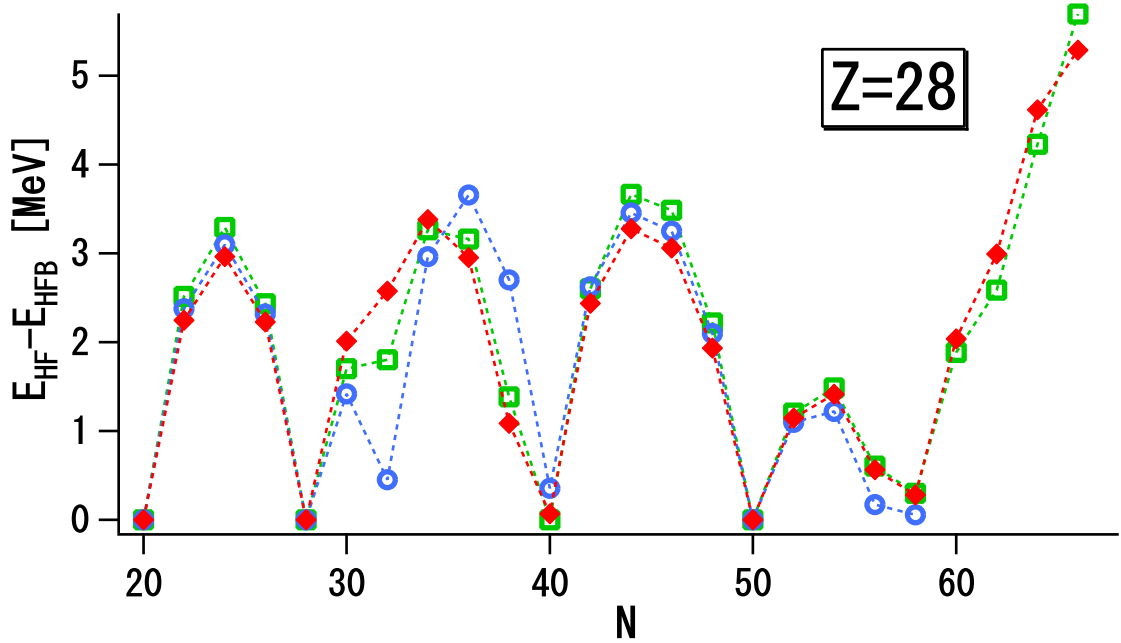}
\caption{Difference between the HF and HFB energies
for the Ni isotopes ($N=\mbox{even}$).
See Fig.~\protect\ref{fig:Sn_Dmass} for conventions.
\label{fig:Z28_Epr}}
\end{figure}

We depict difference between the HF and the HFB energies,
which represents the pair correlation,
for the Ca nuclei in Fig.~\ref{fig:Z20_Epr}
and for the Ni isotopes in Fig.~\ref{fig:Z28_Epr}.
These results are qualitatively similar to those shown
in Ref.~\cite{ref:Nak08b}.
However, we dare to repeat several interesting points.
The vanishing difference between the HF and HFB energies
is normally connected to the shell closure.
It is found that $^{52}$Ca is nearly a doubly-magic nucleus
with any of the interactions,
as is consistent with the experiments~\cite{ref:Ca52_Ex2}.
With M3Y-P5$'$, $^{68}$Ni is almost doubly magic
in harmony with experimental data~\cite{ref:Ni68},
although $^{60}$Ca is not.
This suggests significant $Z$-dependence of the shell structure
around $N=40$,
whose origin will be argued elsewhere.
The hindrance of the pair excitation at $^{86}$Ni
suggests magic or submagic nature of $N=58$
due to the gap between $n2s_{1/2}$ and $n1d_{3/2}$.

Prediction of the neutron drip line depends on effective interactions
to a certain degree.
Compared to the results with M3Y-P4 and P5~\cite{ref:Nak08b},
more neutron-rich Ca and Ni nuclei are bound with M3Y-P4$'$ and P5$'$.
The heaviest bound Ca (Ni) nucleus has $N=50$ ($66$)
in both of the M3Y-P4$'$ and P5$'$ results.

\section{Summary\label{sec:summary}}

We have obtained new parameter-sets
of the M3Y-type semi-realistic effective interaction
to describe structure of nuclei,
by taking into account the whole contribution of the interaction
to the pairing in the HFB calculations.
Two new parameter-sets have been presented;
one keeping the tensor force of the M3Y-Paris interaction (M3Y-P5$'$),
and the other discarding the tensor force (M3Y-P4$'$).
The OPEP part in the central force is maintained.
Basic characters of the interactions are examined
in the infinite nuclear matter and in several doubly magic nuclei.
The pairing properties are checked by the even-odd mass difference
of the Sn isotopes in the HFB approximation.
Pair energies and location of the neutron drip line
are investigated for the Ca and Ni isotopes.
The results do not change significantly
from the previous ones,
in which influence of the density-dependent contact force
on the pairing was partly discarded.

\begin{acknowledgments}
This work is financially supported
as Grant-in-Aid for Scientific Research (C), No.~19540262,
by Japan Society for the Promotion of Science.
Numerical calculations are performed on HITAC SR11000
at Institute of Media and Information Technology, Chiba University,
at Information Technology Center, University of Tokyo,
and at Information Initiative Center, Hokkaido University.
\end{acknowledgments}


\end{document}